\documentclass{article}

\usepackage{PRIMEarxiv}

\usepackage[utf8]{inputenc} 
\usepackage[T1]{fontenc}    
\usepackage{hyperref}       
\usepackage{url}            
\usepackage{booktabs}       
\usepackage{amsfonts}       
\usepackage{nicefrac}       
\usepackage{microtype}      
\usepackage{lipsum}
\usepackage[numbers]{natbib}
\usepackage{fancyhdr}       
\usepackage{graphicx}       
\usepackage{siunitx}
\usepackage{todonotes}
\usepackage{caption}
\usepackage{subcaption}
\graphicspath{{media/}}     
\usepackage{tikz,xcolor,hyperref}
\definecolor{lime}{HTML}{A6CE39}
\DeclareRobustCommand{\orcidicon}{%
    \begin{tikzpicture}
    \draw[lime, fill=lime] (0,0) 
    circle [radius=0.16] 
    node[white] {{\fontfamily{qag}\selectfont \tiny ID}};    \draw[white, fill=white] (-0.0625,0.095) 
    circle [radius=0.007];    \end{tikzpicture}
    \hspace{-2mm}}
\foreach \x in {A, ..., Z}{%
    \expandafter\xdef\csname orcid\x\endcsname{\noexpand\href{https://orcid.org/\csname orcidauthor\x\endcsname}{\noexpand\orcidicon}}
    }
\title{Spontaneous Vibrations and Stochastic Resonance of Short Oligomeric Springs
}

\author{
  Alexey M. Astakhov$^{1,2,a}$ , Vladislav S. Petrovskii$^{1,2,a}$\orcidF, Maria A. Frolkina$^{1,2}$\orcidD, Anastasia A. Markina$^{1,2}$\orcidA{},\\ 
  \textbf{Alexander D. Muratov$^{1,2,b}$\orcidE, Alexander F. Valov$^{1,2}$ and Vladik A. Avetisov$^{1,2,c}$\orcidB } \\
  $^{1}$ Semenov Federal Research Center for Chemical Physics, Russian Academy of Sciences, Moscow, Russia \\
  $^{2}$ Design Center for Molecular Machines, Moscow, Russia \\
  \texttt{$^{a}$ A.M.A. and V.S.P. contributed equally to this work} \\
  \texttt{$^{b}$ alexander.muratov@chph.ras.ru} \\
  \texttt{$^{c}$ avetisov@chph.ras.ru} \\
}

\begin{document}
\maketitle

\begin{abstract}
There is growing interest in molecular structures that exhibit dynamics similar to bistable mechanical systems. These structures have the potential to be used as two-state operating units for various functional purposes. Particularly intriguing are the bistable systems that display spontaneous vibrations and stochastic resonance. Previously, through molecular dynamics simulations, it was discovered that short pyridine-furan springs in water, when subjected to stretching with power loads, exhibit the bistable dynamics of a Duffing oscillator. In this study, we extend these simulations to include short pyridine-pyrrole and pyridine-furan springs in a hydrophobic solvent. Our findings demonstrate that these systems also display the bistable dynamics, accompanied by spontaneous vibrations and stochastic resonance activated by thermal noise.    
\end{abstract}


\section{Introduction}
The increasing focus on nanoscale molecular structures with dynamics reminiscent of bistable mechanical systems is driven by the growing demand for designing and implementing various nanodevices. These devices serve as switches and logic gates\cite{mi6081046,C5SC02317C,C7CS00491E,benda_substrate-dependent_2019,berselli_robust_2021,NICOLI2021213589}, sensors and actuators\cite{zhang_molecular_2018,shu_stimuli-responsive_2020,lemme_nanoelectromechanical_2020,shi_driving_2020,aprahamian_future_2020}, mechanoelectrical transducers and energy harvesters\cite{li_energy_2014,kim_harvesting_2015,ackerman_anomalous_2016,dutreix_two-level_2020,thibado_fluctuation-induced_2020}. Nanoscale bistable systems are equally crucial for validating the principles of stochastic thermodynamics\cite{evans_fluctuation_2002,seifert_stochastic_2012,horowitz_thermodynamic_2020,ciliberto_experiments_2017}. This field currently aims to expand thermodynamic theories to encompass nanoscale molecular machines\cite{ciliberto_experiments_2017,wang_experimental_2002,jop_work_2008,astumian_stochastic_2018,vroylandt_efficiency_2020}.

This article focuses on nanoscale molecular structures whose dynamic prototypes are the textbook bistable mechanical systems known as Euler arches\cite{arnold_catastrophe_1984,poston_catastrophe_1996} and Duffing oscillators\cite{duffing1918erzwungene,korsch_duffing_2008}. In particular, using molecular dynamics simulations, it was found that the long-term conformational dynamics of short rod-shaped thermosensitive oligomers were similar to the bistable dynamics of an Euler arch\cite{avetisov2019oligomeric,markina2020detection}, while particular oligomers of a helical form stabilized by weak intermolecular interactions could behave as bistable Duffing oscillators\cite{avetisov2021short}. 

In general, Euler arches and Duffing oscillators can be perceived as mechanical systems whose potential energy is determined by fourth-degree polynomial function caused with nonlinear elasticity of the system elements. The potential energy of these systems has either one minimum or two energy wells separated by a bistability barrier. Therefore, the system can have different dynamic modes  controlled by the power loads applied to the system. In particular, the force load compressing an Euler arch controls the arch dynamics, while the stretching of a Duffing spring controls the spring dynamics. Therefore, with driving the power loads, one can operate by  dynamical modes of the system and, accordingly, carry out the transitions between the system discrete states in a controlled manner.

In addition to deterministic transitions, jump-like spontaneous transitions between the two states of a bistable system, known as spontaneous vibrations, can occur due to random disturbances of the system. In the mode of spontaneous vibrations, the time intervals between the jumping (the lifetimes of the system in its states) are random values whose average  grows exponentially with increasing ratio of the bistability barrier to the intensity of noise, following Kramer’s rate approximation\cite{KRAMERS1940284}. Therefore, spontaneous vibrations are actually observed when this ratio is not excessively large, such as when the bistability barrier is approximately an order of magnitude higher than the noise intensity. On the other hand, by applying a gentle oscillating force that rocks the bistable potential, spontaneous vibrations can be transformed into almost regular switching between the two states, induced by noise. This phenomenon is referred to as stochastic resonance\cite{benzi_mechanism_1981}. Both spontaneous vibrations and stochastic resonance are remarkable manifestations of bistability.

Stochastic resonance is an intriguing phenomenon that arises from the interplay between the bistable dynamics of a system and its stochastic perturbations. Unlike typical noise effects that tend to blur signals, the noise actually amplifies weak signals in the stochastic resonance regime. While the initial concept of stochastic resonance was proposed to explain the regularity of ice ages on Earth\cite{benzi_mechanism_1981,benzi_stochastic_1982,benzi_theory_1983}, it sparked a proliferation of research exploring its practical applications and interpretations in various macroscopic, global, and even celestial systems\cite{gammaitoni_stochastic_1998,wellens_stochastic_2004}.

In recent years, experimental evidence has emerged indicating the potential presence of bistable patterns at sub-micron scales, such as in nanotubes\cite{baughman1999carbon,fujii2017single,huang_nonlocal_2019}, graphene sheets\cite{ackerman_anomalous_2016,liang2012electromechanical}, DNA hairpins, and proteins\cite{forns2011improving,hayashi2012single,cecconi2005direct}. It is worth noting that spontaneous vibrations and stochastic resonance in macroscopic mechanical systems, even those as small as a micron, are unlikely to be triggered solely by environmental thermal noise. The bistability barriers in macroscopic systems are much higher than the intensity of thermal noise ($\sim k_{B}T$) at natural conditions, necessitating much stronger random perturbations to activate spontaneous vibrations and stochastic resonance, even at the micron scale.

Designing a mechanical system with the nonlinear elasticity inherent in a Duffing oscillator is also a non-trivial task. In mechanics, creative combinations of springs have been devised to mimic the Duffing's bistability (for an example, refer to \citet{LAI201660,LU2021249}). However, in nanoscale mechanics, a potential solution may arise primarily due to strong nonlinearity of weak intermolecular interactions whose contribution to the potential energy of a nano-size molecular system may appear significant for collective dynamical modes of the system. An equally important perspective comes from the fact that the bistability barriers of nanoscale systems can be high enough to well separate dynamic states against a background of thermal fluctuations, and at the same time low enough for thermal fluctuations to activate transitions between the well-separated states. A ratio of bistability barrier to thermal noise intensity of approximately ten can serve as a reasonable benchmark. Oligomeric molecules a few nanometers in size appear to be potential representatives of target molecular systems. 

Notably, recent intensive molecular dynamic simulations investigating short oligomeric compounds subjected to force loads have uncovered bistable molecules that exhibit the dynamic behavior resembling that of the Euler arches and Duffing oscillators\cite{avetisov2019oligomeric,markina2020detection,avetisov2021short}. These simulations have revealed the presence of mechanic-like bistability in specific oligomeric molecules, accompanied by spontaneous vibrations and stochastic resonance activated by thermal fluctuations. In this paper we continue our search for bistable nanoscale molecular structures and investigate behaviour of short pyridine-pyrrole (PP) and pyridine-furan (PF) springs in hydrophobic solvent.

\section{Matherials and Methods}
\label{sec:MatMet}

\subsection{Pyridine-Pyrrole and Pyridine-Furan Springs}\label{Sec:MatMet1}
\begin{figure}
    \centering
    \includegraphics[width=\textwidth]{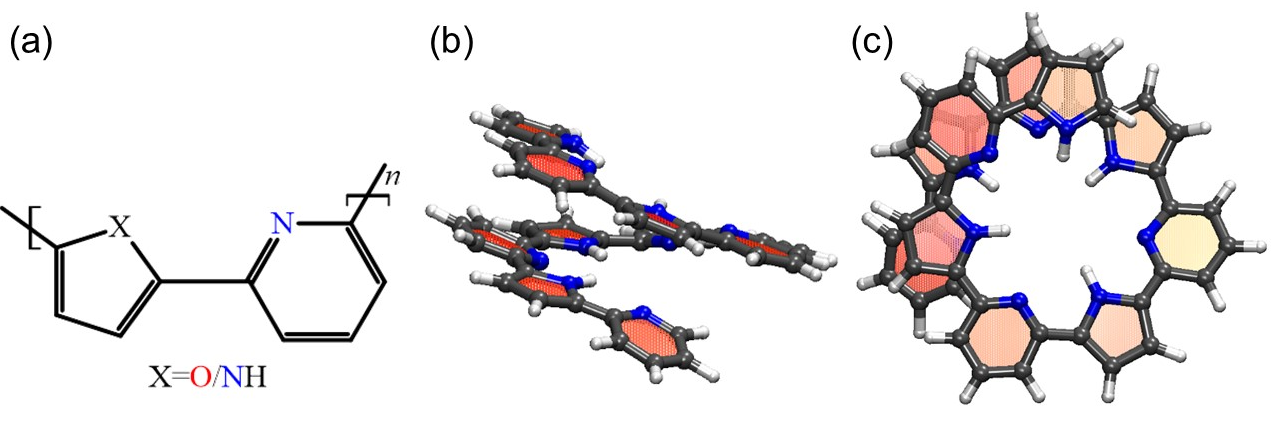}
    \caption{Pyridine–furan (PF) and pyridine-pyrrole (PP) springs with five monomer units (oligo-PF-$5$ spring and oligo-PP-$5$ spring, respectively): (a) Chemical structure of a pyridine-furan ($X$ represents oxygen, $O$) /pyridine-pyrrole monomer ($X$ represents $NH$ group) unit with heterocyclic rings in cis-configuration. (b) Front and (c) top views of an oligo-PP-$5$ spring in the atomistic representation. The spring has one complete turn consisting of approximately $3.5$ monomer units.}
    \label{fig:fig1}
\end{figure}
PP and PF copolymers (Figure \ref{fig:fig1}a) are conductive polymers consisting of $5$- and $6$-member heterocyclic rings as synthesized and described by \citet{ALANJONES19968707} and \citet{jones1997extended}, respectively. These copolymers tend to assume a helix-like shape, which is squeezed by the $\pi -\pi$ interactions of aromatic groups located at the adjacent turns\cite{sahu2015}. Assuming that stacking could lead to nonlinear elasticity of the springs and following the quantum calculations of the stacking energy for different configurations of heterocyclic rings\cite{sahu2015}, the cis-configuration of oligo-PP and oligo-PF with heteroatoms of the $5$- and $6$-member heterocyclic rings on one side of a polymer chain was selected (see Figure \ref{fig:fig1}a)). Guided by the preliminary screening of sizes, we designed two spring models consisting of five monomer units (oligo-PP-$5$ and oligo-PF-$5$) as shown in Figure \ref{fig:fig1}b,c). Both springs were solvated in hydrophobic solvent, tetrahydrofuran (THF). The distance between the adjacent turns was close to $\SI{0.35}{\nano\metre}$ in all non-stretched samples according to \citet{sahu2015}.

\subsection{Simulation Details}\label{Sec:MatMet2}

The oligo-PP- and oligo-PF-springs and the environmental solvent were modeled in a fully atomistic representation with a canonical (symbol/volume/temperature [NVT]) ensemble (box size: $\SI{4.5}{} \times \SI{4.5}{} \times \SI{4.5}{\nano\metre\cubed}$ for oligo-PP, $\SI{7.0}{} \times \SI{7.0}{} \times \SI{7.0}{\nano\metre\cubed}$ for oligo-PF) with a time step of $\SI{2}{\femto\second}$ using Gromacs $2019$\cite{abr2015} and the OPLS-AA\cite{kam2001} force field parameters (for more details, see Parameters for Molecular Dynamics simulation section of Supporting Information
). The temperature was set at $\SI{280}{\kelvin}$ by the velocity-rescale thermostat\cite{bussi2007canonical}, which corresponds to the equilibrium state of the springs\cite{sahu2015}. Each dynamic trajectory was $\SI{300}{} - \SI{350}{\nano\second}$ long and was repeated three times to obtain better statistics; therefore. the effective length of the trajectories was about one $\SI{}{\micro\second}$ for each sample.

The dynamics of the springs were studied by fixing one end of the spring, while the other end was pulled by a force applied along the axis of the spring. The distance (denoted  $R_{e}$) between the ends of the spring (yellow and blue balls in Figure \ref{fig:fig2}a)) was considered a collective variable describing the long-term dynamics of the spring. Bistability of the spring was specified in agreement with two well reproduced states of the spring with the end-to-end distances equal to $R_{e} \sim \SI{1.10}{\nano\metre}$ and  $R_{e} \sim \SI{1.45}{\nano\metre}$. These states are referred to as the squeezed and the stress-strain states, respectively.

\section{Results}
\label{sec:Res}
\subsection{Bistable dynamics of oligo-PP-5 spring}\label{sec:Res1}
To investigate the dynamics of the oligo-PP-$5$ springs under tension, we initially equilibrated the oligo-PP-$5$ spring at $\SI{280}{\kelvin}$ with one end fixed. Subsequently, we applied a force $\vec{F}$ along the spring axis to pull the other end. Under weak tensile conditions, the spring's initial state, compressed by stacking, remained stable, and the spring underwent slight stretching in accordance with linear elasticity. Note that for oligo-PP-$5$ spring we measured not the end-to-end distance, but the distance between the pulled end and the monomer that contacts the pulled end in the squeezed state (see comparison in SI). However, when the pulling force reached a specific critical value of approximately $F_{c}=\SI{30}{\pico\newton}$, the oligo-PP-$5$ spring exhibited bistability and commenced to vibrate spontaneously. At the critical force value, a junction point emerged, leading to a split into two branches: the branch of zero-stress attractors, representing a stress-strain state, and the branch of unsteady zero-stress states, repelling the dynamic trajectories.
\begin{figure}
    \centering
    \includegraphics[width=\textwidth]{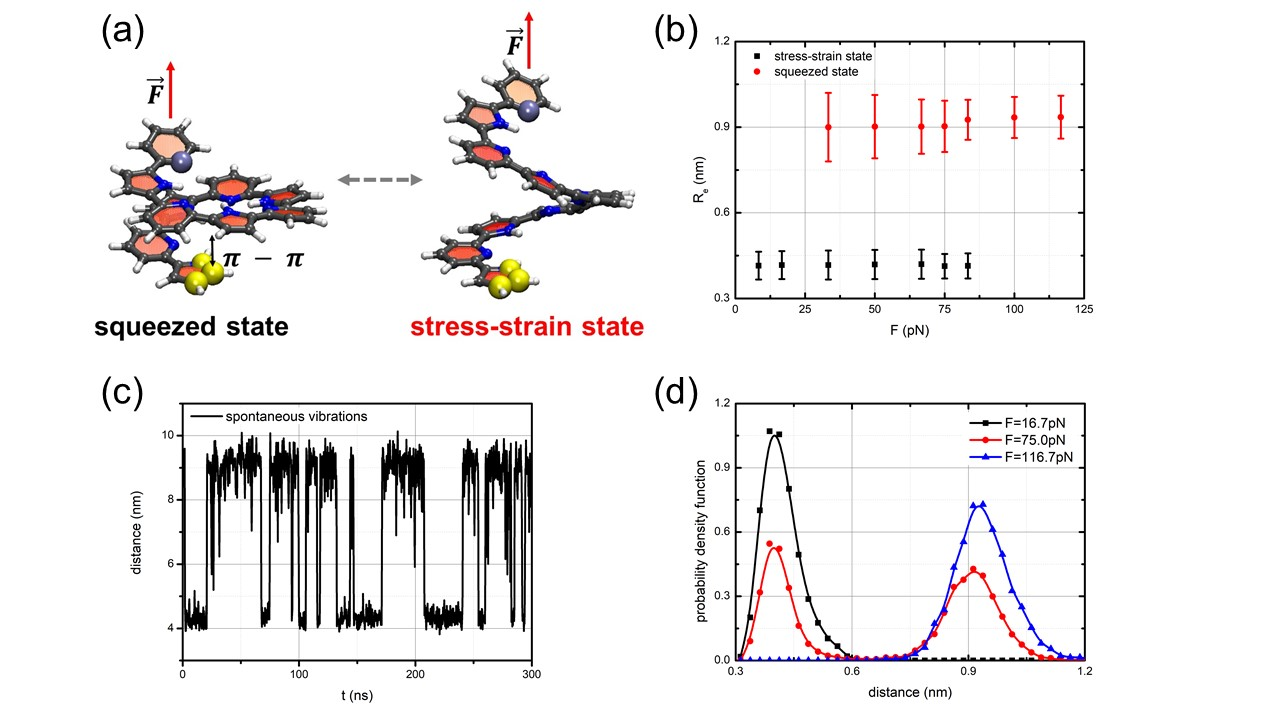}
    \caption{(a) Computational models of the oligo-PP-$5$ system with applied longitudinal load. The squeezed and the stress–strain states of the spring are shown on the right and left, respectively. The pulling force, $F$, is applied to the top end of the spring; (b) The state diagram shows a linear elasticity of oligo-PF-$5$ spring up to  $F\approx \SI{30}{\pico\newton}$ and bistability of the spring in the region from $F\approx\SI{30}{}- \SI{80}{\pico\newton}$; (c) Spontaneous vibrations of the oligo-PP-$5$ spring at  $F\approx \SI{75}{\pico\newton}$; (d) Evolution of the probability density for the squeezed and stress–strain states when pulling force surpasses the critical value.}
    \label{fig:fig2}
\end{figure}

Simultaneously, the squeezed states remain attractive. From the perspective of nonlinear dynamical systems, the dynamics of the oligo-PP-$5$ spring bifurcate at the critical force $F_{c}=\SI{30}{\pico\newton}$. Beyond this critical tensile point, the spring becomes bistable and exhibits spontaneous vibrations, alternating between the squeezed and stress-strain states. The average end-to-end distances of the spring in the squeezed and stress-strain states differ by approximately $\SI{0.35}{\nano\metre}$, allowing for clear distinction between these two states. Figure \ref{fig:fig2}a) shows atom level snapshots of these two states. Notably, this difference suggests an extension of the stacking pair length to almost twice its original size. As a result, the $\pi-\pi$ interactions do not significantly contribute to the elastic energy of the stress-strain states, with the spring's elasticity mainly determined by the rigidity of the oligomeric backbone.

Figure \ref{fig:fig2}d) illustrates the evolution of the statistics of visits to the squeezed and stress-strain states as the pulling force exceeds the critical point $F_{c}$. Below $F_{c}$, the squeezed state represents the sole steady state of the spring. However, at the bifurcation point $F_{c}$, the stress-strain state emerges, rendering the oligo-PP-$5$ spring bistable, causing it to spontaneously vibrate, although the squeezed state predominates near the critical point $F_{c}$. At $F\approx\SI{75}{\pico\newton}$, both the squeezed and stress-strain states are almost equally visited, signifying that the bistability of the oligo-PP-$5$ spring becomes approximately symmetrical at a considerable distance from the critical point.

In this region, the spontaneous vibrations of the oligo-PP-$5$ spring are most pronounced. The mean lifetimes of the squeezed and stress-strain states in the spontaneous vibration mode varied in the bistability region, ranging from $\tau=\SI{1}{}-\SI{20}{\nano\second}$, depending on the pulling force. In the symmetrical bistability region, neither the squeezed state nor the stress-strain state dominates, resulting in the mean lifetimes of the two states being approximately equal to $\tau=\SI{14}{\nano\second}$.

Utilizing Kramer’s rate approximation with a collision time for random perturbations ranging from $0.1-\SI{10}{\pico\second}$, we can roughly estimate the bistability barrier of the oligo-PP-$5$ spring as $\sim 10$ $k_{B}T$. Noteworthy, the bistability barrier of this bistable oligomeric system is approximately ten times greater than the characteristic scale of thermal fluctuations, $k_{B}T$.
\begin{figure}
    \centering
    \includegraphics[width=\textwidth]{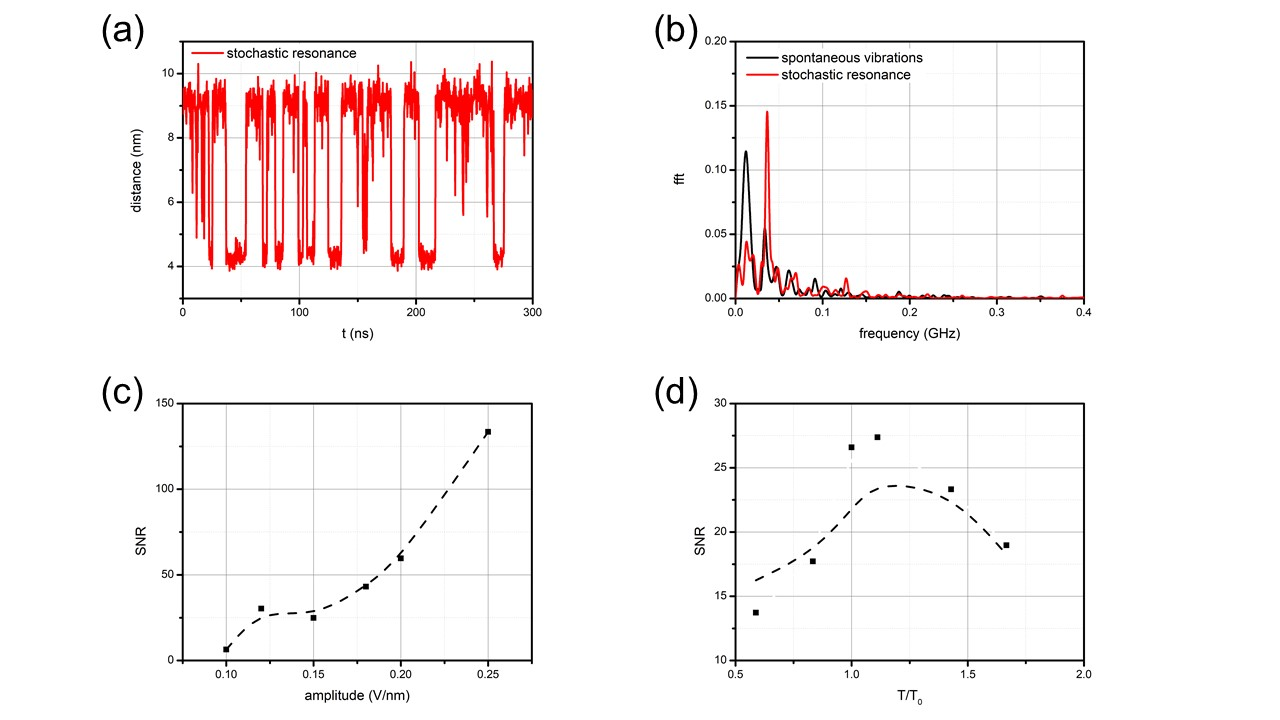}
    \caption{Stochastic resonance of the oligo-PP-$5$ induces by an oscillating field $E = E_{0} \cos (2 \pi \nu t)=E_{0} \cos (\nicefrac{2 \pi t}{T})$: (a) The dynamic trajectory at $F = \SI{3.5}{\pico\newton}$, $T = \SI{28}{\nano\second}$, and $E_{0} = \SI{0.12}{\volt\per\nano\metre}$; (b) Power spectrum of spontaneous vibrations (black curve) and stochastic resonance (red curve); (c) The dependence of the main resonance peak amplitude on $E_{0}$ ($T_{0}=\SI{28}{\nano\second}$); (d) The dependence of the main resonance peak amplitude on the period $T$ of oscillating field ($E_{0}=\SI{0.12}{\volt\per\nano\metre}$).}
    \label{fig:fig3}
\end{figure}
Figure \ref{fig:fig2}c) displays a typical trajectory of the long-term dynamics $R_{e}(t)$ of the oligo-PP-$5$ spring within the symmetric bistability region. Clear spontaneous vibrations of the spring can be observed without any additional random perturbations applied to activate them. Instead, these vibrations are solely activated by thermal-bath fluctuations. On the other hand, outside the bistability region, non-vibrating trajectories are prevalent.

To investigate the stochastic resonance mode of the oligo-PP-$5$ spring, we introduced an additional weak oscillating force applied to the pulling end of the spring. This oscillating force was modeled by applying an oscillating electrical field, $E=E_{0} \cos (2\pi \nu t)$, to a unit charge located on the pulling end of the spring. A counterion was placed at $\SI{2.2}{\nano\metre
}$ from spring centre of mass on the pull axis to balance the system. We need to note that addition of the charge and a fixed counterion significantly changes the critical force parameters of bistability - for this particular configuration, system exhibits spontaneous vibrations even without additional pulling force, and reaches symmetrical distribution at $F = \SI{3.5}{\pico\newton}$
(for more details, refer to the "Parameters of periodic signal" section in the Supporting Information).

Figure \ref{fig:fig3}a) presents typical vibrations of the end-to-end distance of the oligo-PP-$5$ spring in the stochastic resonance mode, along with the power spectrum of these vibrations.

After some preliminary analysis, we decided to examine frequency response at oscillating field amplitude $E_{0}=\SI{0.12}{\volt\per\nano\metre}$, because at amplitudes above $E_{0} > \SI{0.15}{\volt\per\nano\metre}$ system behavior resembled forced oscillations and not stochastic resonance (mean lifetime in state was directly proportional to external signal period). 
According to the theory of stochastic resonance \cite{gammaitoni_stochastic_1998,wellens_stochastic_2004}, the primary resonance peak was observed at a frequency of $\nu=\nicefrac{1}{2\tau}$, where the period of the applied oscillating field was equal to twice the mean lifetime of the states in the spontaneous vibration mode. We extensively scanned a wide range of oscillating fields to identify the maximal resonance response, as determined by the spectral component at the resonance frequency. Figure \ref{fig:fig3}c-d) presents the corresponding results. The maximum resonance response was observed when the period of the oscillating field was close to twice the mean lifetime of the states in the spontaneous vibration mode. 

\subsection{Bistable dynamics of oligo-PF-5 spring}\label{sec:Res2}

\begin{figure}
    \centering
    \includegraphics[width=\textwidth]{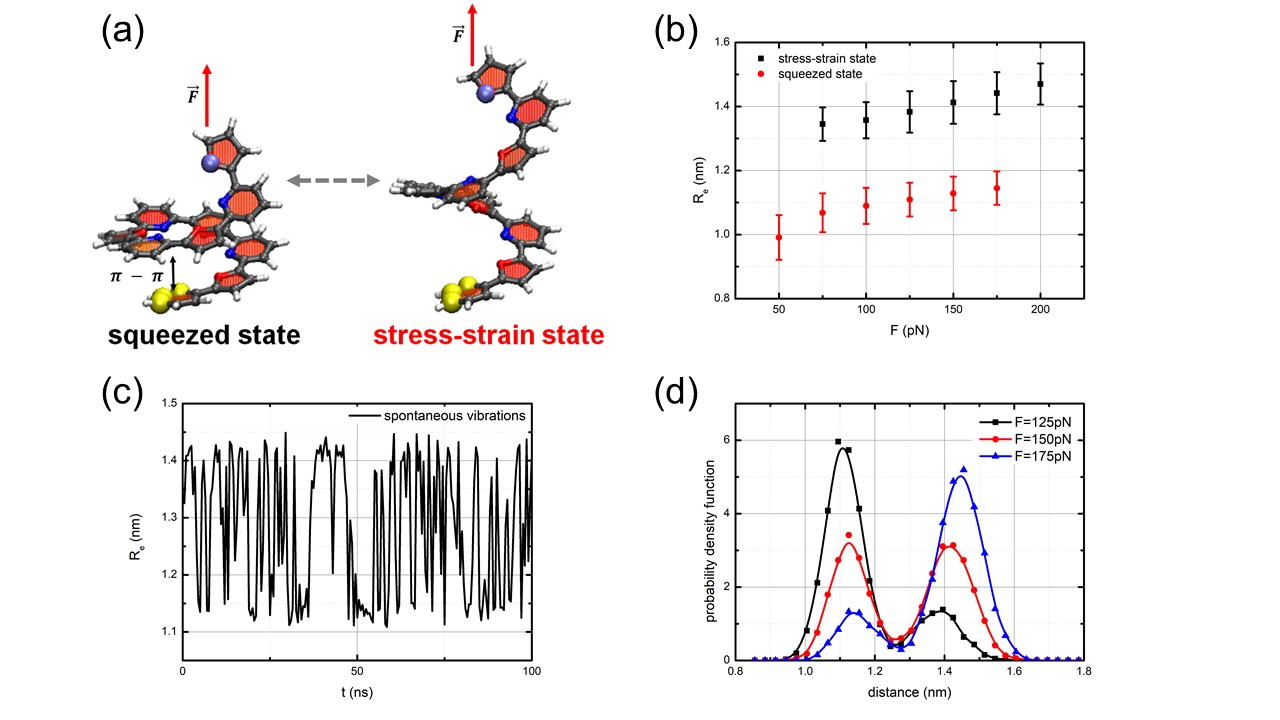}
    \caption{(a) Computational models of the oligo-PF-$5$ system with applied longitudinal load. The squeezed and the stress–strain states of the spring are shown on the left and right, respectively. The yellow spheres at the lower end of the spring indicate the fixation of the pyridine ring by rigid harmonic force. The pulling force, $F$, is applied to the top end of the spring. (b) The state diagram shows a linear elasticity of oligo-PF-$5$ spring up to  $F\approx \SI{50}{\pico\newton}$ and bistability of the spring in the region from $F\approx\SI{50}{}- \SI{200}{\pico\newton}$; (c) Spontaneous vibrations of the oligo-PF-$5$ spring at  $F\approx \SI{150}{\pico\newton}$; (d) Evolution of the probability density for the squeezed and stress–strain states when pulling force surpasses the critical value.}
    \label{fig:fig4}
\end{figure}

To investigate the dynamics of the oligo-PF-$5$ springs under tension, we followed the same protocol as for the oligo-PP-$5$. First, we equilibrated the oligo-PF-$5$ spring at $\SI{280}{\kelvin}$ with one end fixed. Then, we applied a force $\vec{F}$ along the spring axis to pull the other end. Once again, under weak tensile conditions, the spring's initial state remained stable, and it underwent slight stretching in line with linear elasticity due to the stacking. However, once the pulling force reached a specific critical value, the oligo-PF-$5$ spring exhibited bistability and began to exhibit spontaneous vibrations in the same way as the oligo-PP-$5$. As the pulling force reached the critical value, approximately $F_{c}=\SI{50}{\pico\newton}$, a junction point emerged, dividing the system into two branches: one corresponding to the stress–strain state with zero-stress attractors, and the other representing unsteady zero-stress states that repelled dynamic trajectories. Simultaneously, the squeezed states remain attractive. The average end-to-end distances of the spring in the squeezed and stress-strain states differ by approximately $\SI{0.35}{\nano\metre}$, allowing for clear distinction between these two states. Figure \ref{fig:fig4}a) displays atomic level snapshots of these two states. Notably, this difference suggests an extension of the stacking pair length to almost twice its original size. As a result, the $\pi-\pi$ interactions do not significantly contribute to the elastic energy of the stress-strain states, with the spring's elasticity mainly determined by the rigidity of the oligomeric backbone.

Figures \ref{fig:fig4}b) and d) illustrate the evolution of the statistics of visits to the squeezed and stress-strain states as the pulling force exceeds the critical point $F_{c}$. Below $F_{c}$, the squeezed state represents the sole steady state of the spring. However, at the bifurcation point $F_{c}$, the stress-strain state emerges, rendering the oligo-PF-$5$ spring bistable, causing it to spontaneously vibrate, although the squeezed state predominates near the critical point $F_{c}$. Within the range of $F=\SI{125}{} - \SI{175}{\pico\newton}$, both the squeezed and stress-strain states are almost equally visited, signifying that the bistability of the oligo-PF-$5$ spring becomes approximately symmetrical at a considerable distance from the critical point.

In this region, the spontaneous vibrations of the oligo-PF-$5$ spring are most prominent. The mean lifetimes of the squeezed and stress-strain states in the spontaneous vibration mode differed in the bistability region, ranging from $\tau=\SI{1}{\nano\second}-\SI{3}{\nano\second}$, depending on the pulling force. In the symmetrical bistability region, neither the squeezed state nor the stress-strain state dominates, resulting in the mean lifetimes of the two states being approximately equal to $\tau=\SI{2.04}{\nano\second}$. Utilizing Kramer’s rate approximation with a collision time for random perturbations ranging from $0.1-\SI{10}{\pico\second}$, we can roughly estimate the bistability barrier of the oligo-PF-$5$ spring as $10$ $k_{B}T$. Again, the bistability barrier of this bistable oligomeric system is approximately ten times greater than the characteristic scale of thermal fluctuations, $k_{B}T$.

\begin{figure}
    \centering
    \includegraphics[width=\textwidth]{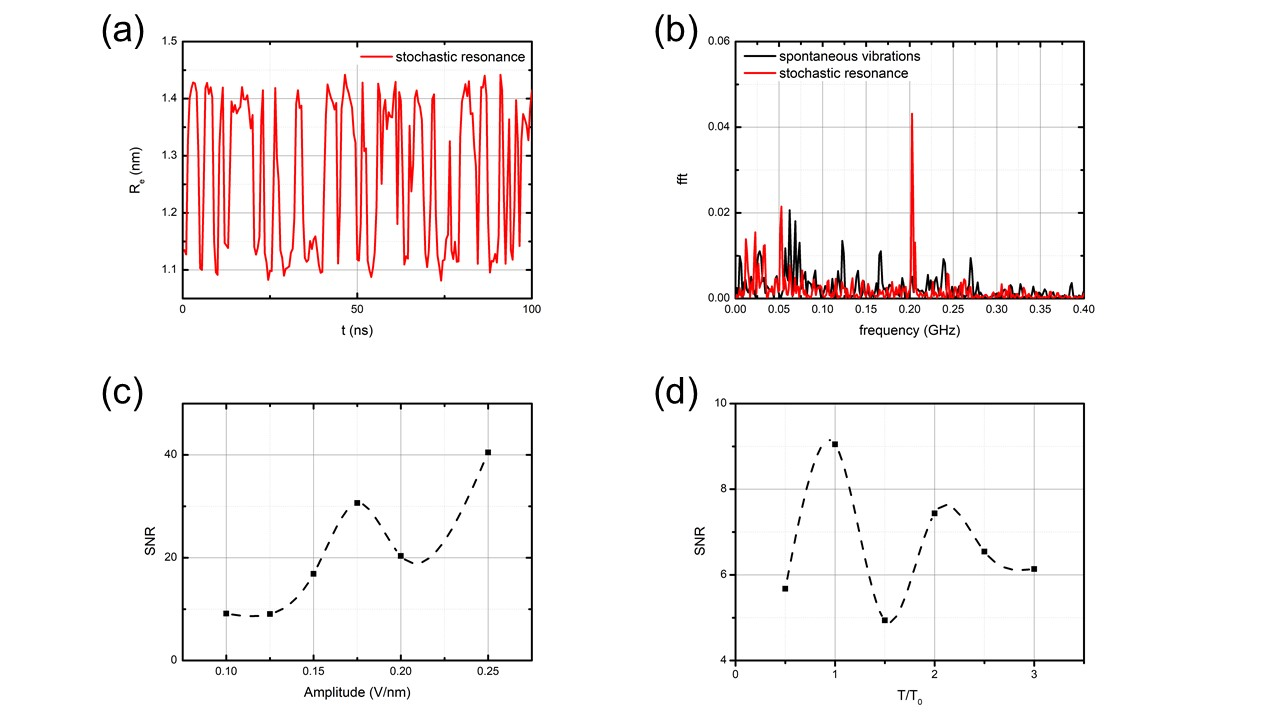}
    \caption{Stochastic resonance of the oligo-PF-5 induces by an oscillating field $E = E_{0} \cos (2 \pi \nu t)=E_{0} \cos (\nicefrac{2 \pi t}{T})$: (a) The dynamic trajectory at $F = \SI{15}{\pico\newton}$, $T = \SI{4.9}{\nano\second}$, and $E_{0} = \SI{0.175}{\volt\per\nano\metre}$; (b) Power spectrum of spontaneous vibrations (red curve) and stochastic resonance (black curve); (c) The dependence of the main resonance peak amplitude on $E_{0}$ ($T_{0}=\SI{4.9}{\nano\second}$); (d) The dependence of the main resonance peak amplitude on the period $T$ of oscillating field ($E_{0}=\SI{0.175}{\volt\per\nano\metre}$).}
    \label{fig:fig5}
\end{figure}

Figure \ref{fig:fig4}c) displays a typical trajectory of the long-term dynamics $R_{e}(t)$ of the oligo-PF-$5$ spring within the symmetric bistability region. Clear spontaneous vibrations of the spring can be observed without any additional random perturbations applied to activate them. Instead, these vibrations are solely activated by thermal-bath fluctuations. On the other hand, outside the bistability region, non-vibrating trajectories are prevalent.

To investigate the stochastic resonance mode of the oligo-PF-$5$ spring, we introduced an additional weak oscillating force by applying an oscillating electrical field, $E=E_{0} \cos (2\pi \nu t)$, to a unit charge located on the pulling end of the spring. A compensative charge was placed on the fixed end to balance the system (for more details, refer to the "Parameters of periodic signal" section in the Supporting Information). Similar to the oligo-PP-$5$ spring, the adding of a charge and
a fixed counterion significantly shifts the bistability region - for this particular configuration,
system exhibits spontaneous vibrations with symmetrical distribution at $F = \SI{15}{\pico\newton}$. Figure \ref{fig:fig5}a) presents typical vibrations of the end-to-end distance of the oligo-PF-$5$ spring in the stochastic resonance mode, along with the power spectrum of these vibrations.

The primary resonance peak was observed at a frequency of $\nu=\nicefrac{1}{2\tau}$, where the period of the applied oscillating field was equal to twice the mean lifetime of the states in the spontaneous vibration mode. We extensively scanned a wide range of oscillating fields to identify the maximal resonance response, as determined by the spectral component at the resonance frequency. Figure \ref{fig:fig5}c-d) presents the corresponding results. The maximum resonance response was observed when the period of the oscillating field was close to twice the mean lifetime of the states in the spontaneous vibration mode. In terms of the amplitude of the oscillating field, the maximum resonance occurred at $E_{0}=\SI{0.175}{\volt\per\nano\metre}$. Notably, the resonance response was diminished in the symmetric bistability region at $F=\SI{15}{\pico\newton}$. Beyond this region, the lifetimes of the squeezed and stress-strain states became substantially different, rendering the average lifetime less indicative of the resonance frequency.

\section{Discussion}\label{sec:Dis}
The main finding of this work is that spontaneous vibrations and stochastic resonance are not exclusive for oligo-PF springs in water\cite{avetisov2021short}, but are present in various compounds and observed in different solvents. However, these bistability effects might vary for different systems; such variations are discussed below.

First note concerns the oligo-PF-$5$ solvated in tetrahydrofuran. In such system, both spontaneous vibrations and stochastic resonance occur at lesser forces than in water. While in case of water the sping's bistability is observed in the region from $F\approx\SI{240}{}-\SI{320}{\pico\newton}$ and the squeezed and the stress-strain states were almost equally visited in the region from $F=\SI{270}{}-\SI{290}{\pico\newton}$, in THF the oligo-PF-$5$ exhibits spontaneous vibrations in the region from $F\approx\SI{50}{}-\SI{200}{\pico\newton}$. Symmetric bistability is achieved at $F=\SI{150}{\pico\newton}$. If we assume that stacking interaction between the turn and the fixed end of the oligo-PF-$5$ spring governs its elasticity at low tensiles while at high tensiles the elasticity imposed be oligomeric backbone becomes dominant, then we must suppose that the $\pi$-stacking is destroyed later in water. Such difference can be explained by the fact that oligo-PF springs are hydrophobic and prefer to remain in squeezed state in hydrophilic environment.

Second note concerns the overstretching limit of the oligo-PF-$5$. Our previous findings show that the overstretching of the oligo-PF-$5$ springs in THF or vacuum occurs at $F=\SI{275}{\pico\newton}$ while in water it happens at $F=\SI{330}{\pico\newton}$(see \citeauthor{avetisov2021short}, Figure S3 of the Supplementary Materials). Moreover, at $F=\SI{330}{\pico\newton}$ all three states, the squeezed, the stress-strain and the overstretched, coexist, so one might assume that the bistability region extends up to the overstretching limit. However, our results of modeling of the oligo-PF-$5$ in THF do not support this idea: the bistability ceases at $F=\SI{200}{\pico\newton}$ and at larger forces the spring exists at the stress-strain state up to the overstretching limit.

Next note concerns the bistability of the oligo-PP-$5$. For this system we observe spontaneous vibrations in the region from $F\approx\SI{30}{}-\SI{80}{\pico\newton}$ and the symmetric bistability is achieved at $F=\SI{75}{\pico\newton}$. Again, above the force of $F=\SI{80}{\pico\newton}$ the oligo-PP-$5$ spring exists in stress-strain state up to overstretching limit, which was $F=\SI{305}{\pico\newton}$ in this case. Interestingly enough, we could not detect the bistable behaviour of the oligo-PP-$5$ in water (See "Oligo-PP-$5$ in water" section of the Supporting Information). This might be due to the fact that although the bistable behavior of the oligo-PP-$5$ might be theoretically expected, the overstretching occurred earlier than the transition from its squeezed state to the stress-strain state since these two phenomena are independent from each other. These three observations once again support the idea that the interplay between short-ranged ($\pi$-stacking, for instance) and long-ranged couplings is crucial for the observation of the bistable behaviour. Fine tuning of the setup might be required whether one desires to reproduce these results experimentally.

The last note concerns the stochastic resonance of both oligo-PF-$5$ and oligo-PP-$5$ springs. If an external oscillating force is strong enough, the forced oscillations of the spring might also happen. These forced oscillations have different origin than the stochastic resonance, the frequency of which is governed by the lifetime of the states in the spontaneous vibrations regime. It is important to distinguish between forced oscillations occurring at large amplitudes of the oscillating field and stochastic resonance observed at weak amplitudes. Based on out results, we determine $E_{0}=\SI{0.175}{\volt\per\nano\metre}$ as the limit of the amplitude; below this limit the stochastic resonance was established.

\section{Conclusion}\label{sec:Con}
We conducted atomic level simulations on short PP- and PF-springs under stretching conditions, and the results revealed clear indications of bistable dynamics characteristic of Duffing oscillators. During the study, we fixed one end of the springs while pulling the other end along the spring's axis. We observed typical bistability characteristics, such as spontaneous vibrations and stochastic resonance, in both springs. To explore the symmetrical bistability conditions thoroughly, we examined a wide range of controlling parameters and determined the mean lifetime of the states in the spontaneous vibration mode for each spring.

By using Kramer's rate approximation with collision times ranging from $\SI{0.1}{}$ to $\SI{20}{\pico\second}$, we estimated the bistability barriers for both springs to be within the range of $5$ to $15 k_{B}T$. Remarkably, the time scales of spontaneous vibrations and the bistability barriers for the oligo-PP-$5$ and oligo-PF-$5$ springs were found to be similar to those of the oligomeric Duffing oscillator and oligomeric Euler arch, as described in previous studies\cite{avetisov2019oligomeric,markina2020detection,avetisov2021short}. The high bistability barriers of these short oligomeric springs effectively prevent separation of the two states due to thermal noise. However, at the same time, these barriers allow transitions between the states to be activated by thermally enriched fluctuations with higher energy.

Based on our modeling of short PP- and PF-springs, along with the previous modeling of oligomeric Euler arches, we propose that nano-sized oligomeric structures stabilized by short-range, low-energy couplings (e.g., weak hydrogen bonds, hydrophilic-hydrophobic interactions, and $\pi-\pi$ interactions) can indeed exhibit bistability, accompanied by thermally-activated spontaneous vibrations and stochastic resonance.

\section*{Funding}
The authors acknowledge the financial support of the Design Center for Molecular Machines, Moscow, Russia, under the Business Contract $28$/$11$-X$\Phi$-$28.11.2022$ with N. N. Semenov Federal Research Center for Chemical Physics, Russian Academy of Sciences.
\section*{Acknowledgments}
We thank Vladimir Bochenkov, Dmitry Pergushov and Anna Popinako for helpful discussions.


\begin{thebibliography}{52}
\providecommand{\natexlab}[1]{#1}
\providecommand{\url}[1]{\texttt{#1}}
\expandafter\ifx\csname urlstyle\endcsname\relax
  \providecommand{\doi}[1]{doi: #1}\else
  \providecommand{\doi}{doi: \begingroup \urlstyle{rm}\Url}\fi

\bibitem[Peschot et~al.(2015)Peschot, Qian, and Liu]{mi6081046}
Alexis Peschot, Chuang Qian, and Tsu-Jae~King Liu.
\newblock Nanoelectromechanical switches for low-power digital computing.
\newblock \emph{Micromachines}, 6\penalty0 (8):\penalty0 1046--1065, 2015.

\bibitem[Varghese et~al.(2015)Varghese, Elemans, Rowan, and Nolte]{C5SC02317C}
Shaji Varghese, Johannes A. A.~W. Elemans, Alan~E. Rowan, and Roeland J.~M.
  Nolte.
\newblock Molecular computing: paths to chemical turing machines.
\newblock \emph{Chem. Sci.}, 6:\penalty0 6050--6058, 2015.

\bibitem[Erbas-Cakmak et~al.(2018)Erbas-Cakmak, Kolemen, Sedgwick,
  Gunnlaugsson, James, Yoon, and Akkaya]{C7CS00491E}
Sundus Erbas-Cakmak, Safacan Kolemen, Adam~C. Sedgwick, Thorfinnur
  Gunnlaugsson, Tony~D. James, Juyoung Yoon, and Engin~U. Akkaya.
\newblock Molecular logic gates: the past{,} present and future.
\newblock \emph{Chem. Soc. Rev.}, 47:\penalty0 2228--2248, 2018.

\bibitem[Benda et~al.(2019)Benda, Doistau, Rossi-Gendron, Chamoreau,
  Hasenknopf, and Vives]{benda_substrate-dependent_2019}
Lorien Benda, Benjamin Doistau, Caroline Rossi-Gendron, Lise-Marie Chamoreau,
  Bernold Hasenknopf, and Guillaume Vives.
\newblock Substrate-dependent allosteric regulation by switchable catalytic
  molecular tweezers.
\newblock \emph{Communications Chemistry}, 2\penalty0 (1):\penalty0 1--11,
  2019.

\bibitem[Berselli et~al.(2021)Berselli, Gimenez, O’Connor, and
  Keyes]{berselli_robust_2021}
Guilherme~B. Berselli, Aurélien~V. Gimenez, Alexandra O’Connor, and Tia~E.
  Keyes.
\newblock Robust {Photoelectric} {Biomolecular} {Switch} at a
  {Microcavity}-{Supported} {Lipid} {Bilayer}.
\newblock \emph{ACS Applied Materials \& Interfaces}, 13\penalty0
  (24):\penalty0 29158--29169, 2021.

\bibitem[Nicoli et~al.(2021)Nicoli, Paltrinieri, {Tranfić Bakić}, Baroncini,
  Silvi, and Credi]{NICOLI2021213589}
Federico Nicoli, Erica Paltrinieri, Marina {Tranfić Bakić}, Massimo
  Baroncini, Serena Silvi, and Alberto Credi.
\newblock Binary logic operations with artificial molecular machines.
\newblock \emph{Coordination Chemistry Reviews}, 428:\penalty0 213589, 2021.

\bibitem[Zhang et~al.(2018)Zhang, Marcos, and Leigh]{zhang_molecular_2018}
Liang Zhang, Vanesa Marcos, and David~A. Leigh.
\newblock Molecular machines with bio-inspired mechanisms.
\newblock \emph{Proceedings of the National Academy of Sciences}, 115\penalty0
  (38):\penalty0 9397--9404, 2018.

\bibitem[Shu et~al.(2020)Shu, Shen, Zhang, and
  Serpe]{shu_stimuli-responsive_2020}
Tong Shu, Qiming Shen, Xueji Zhang, and Michael~J. Serpe.
\newblock Stimuli-responsive polymer/nanomaterial hybrids for sensing
  applications.
\newblock \emph{Analyst}, 145\penalty0 (17):\penalty0 5713--5724, 2020.

\bibitem[Lemme et~al.(2020)Lemme, Wagner, Lee, Fan, Verbiest, Wittmann, Lukas,
  Dolleman, Niklaus, van~der Zant, Duesberg, and
  Steeneken]{lemme_nanoelectromechanical_2020}
Max~C. Lemme, Stefan Wagner, Kangho Lee, Xuge Fan, Gerard~J. Verbiest,
  Sebastian Wittmann, Sebastian Lukas, Robin~J. Dolleman, Frank Niklaus, Herre
  S.~J. van~der Zant, Georg~S. Duesberg, and Peter~G. Steeneken.
\newblock Nanoelectromechanical {Sensors} {Based} on {Suspended} {2D}
  {Materials}.
\newblock \emph{Research}, 2020, 2020.

\bibitem[Shi et~al.(2020)Shi, Zhang, Tian, and Qu]{shi_driving_2020}
Zhao-Tao Shi, Qi~Zhang, He~Tian, and Da-Hui Qu.
\newblock Driving {Smart} {Molecular} {Systems} by {Artificial} {Molecular}
  {Machines}.
\newblock \emph{Advanced Intelligent Systems}, 2\penalty0 (5):\penalty0
  1900169, 2020.

\bibitem[Aprahamian(2020)]{aprahamian_future_2020}
Ivan Aprahamian.
\newblock The {Future} of {Molecular} {Machines}.
\newblock \emph{ACS Central Science}, 6\penalty0 (3):\penalty0 347--358, 2020.

\bibitem[Li et~al.(2014)Li, Tian, and Deng]{li_energy_2014}
Huidong Li, Chuan Tian, and Z.~Daniel Deng.
\newblock Energy harvesting from low frequency applications using piezoelectric
  materials.
\newblock \emph{Applied Physics Reviews}, 1\penalty0 (4):\penalty0 041301,
  2014.

\bibitem[Kim et~al.(2015)Kim, Lima, Kozlov, Haines, Spinks, Aziz, Choi, Sim,
  Wang, Lu, Qian, Madden, Baughman, and Kim]{kim_harvesting_2015}
Shi~Hyeong Kim, Márcio~D. Lima, Mikhail~E. Kozlov, Carter~S. Haines,
  Geoffrey~M. Spinks, Shazed Aziz, Changsoon Choi, Hyeon~Jun Sim, Xuemin Wang,
  Hongbing Lu, Dong Qian, John D.~W. Madden, Ray~H. Baughman, and Seon~Jeong
  Kim.
\newblock Harvesting temperature fluctuations as electrical energy using
  torsional and tensile polymer muscles.
\newblock \emph{Energy \& Environmental Science}, 8\penalty0 (11):\penalty0
  3336--3344, 2015.

\bibitem[Ackerman et~al.(2016)Ackerman, Kumar, Neek-Amal, Thibado, Peeters, and
  Singh]{ackerman_anomalous_2016}
M.~L. Ackerman, P.~Kumar, M.~Neek-Amal, P.~M. Thibado, F.~M. Peeters, and
  Surendra Singh.
\newblock Anomalous {Dynamical} {Behavior} of {Freestanding} {Graphene}
  {Membranes}.
\newblock \emph{Physical Review Letters}, 117\penalty0 (12):\penalty0 126801,
  2016.

\bibitem[Dutreix et~al.(2020)Dutreix, Avriller, Lounis, and
  Pistolesi]{dutreix_two-level_2020}
C.~Dutreix, R.~Avriller, B.~Lounis, and F.~Pistolesi.
\newblock Two-level system as topological actuator for nanomechanical modes.
\newblock \emph{Physical Review Research}, 2\penalty0 (2):\penalty0 023268,
  2020.

\bibitem[Thibado et~al.(2020)Thibado, Kumar, Singh, Ruiz-Garcia, Lasanta, and
  Bonilla]{thibado_fluctuation-induced_2020}
P.~M. Thibado, P.~Kumar, Surendra Singh, M.~Ruiz-Garcia, A.~Lasanta, and L.~L.
  Bonilla.
\newblock Fluctuation-induced current from freestanding graphene.
\newblock \emph{Physical Review E}, 102\penalty0 (4):\penalty0 042101, 2020.

\bibitem[Evans and Searles(2002)]{evans_fluctuation_2002}
Denis~J. Evans and Debra~J. Searles.
\newblock The {Fluctuation} {Theorem}.
\newblock \emph{Advances in Physics}, 51\penalty0 (7):\penalty0 1529--1585,
  2002.

\bibitem[Seifert(2012)]{seifert_stochastic_2012}
Udo Seifert.
\newblock Stochastic thermodynamics, fluctuation theorems and molecular
  machines.
\newblock \emph{Reports on Progress in Physics}, 75\penalty0 (12):\penalty0
  126001, 2012.

\bibitem[Horowitz and Gingrich(2020)]{horowitz_thermodynamic_2020}
Jordan~M. Horowitz and Todd~R. Gingrich.
\newblock Thermodynamic uncertainty relations constrain non-equilibrium
  fluctuations.
\newblock \emph{Nature Physics}, 16\penalty0 (1):\penalty0 15--20, 2020.

\bibitem[Ciliberto(2017)]{ciliberto_experiments_2017}
S.~Ciliberto.
\newblock Experiments in {Stochastic} {Thermodynamics}: {Short} {History} and
  {Perspectives}.
\newblock \emph{Physical Review X}, 7:\penalty0 021051, 2017.

\bibitem[Wang et~al.(2002)Wang, Sevick, Mittag, Searles, and
  Evans]{wang_experimental_2002}
G.~M. Wang, E.~M. Sevick, Emil Mittag, Debra~J. Searles, and Denis~J. Evans.
\newblock Experimental {Demonstration} of {Violations} of the {Second} {Law} of
  {Thermodynamics} for {Small} {Systems} and {Short} {Time} {Scales}.
\newblock \emph{Physical Review Letters}, 89\penalty0 (5):\penalty0 050601,
  2002.

\bibitem[Jop et~al.(2008)Jop, Petrosyan, and Ciliberto]{jop_work_2008}
P.~Jop, A.~Petrosyan, and S.~Ciliberto.
\newblock Work and dissipation fluctuations near the stochastic resonance of a
  colloidal particle.
\newblock \emph{EPL (Europhysics Letters)}, 81\penalty0 (5):\penalty0 50005,
  2008.

\bibitem[Astumian(2018)]{astumian_stochastic_2018}
R.~D. Astumian.
\newblock Stochastic pumping of non-equilibrium steady-states: how molecules
  adapt to a fluctuating environment.
\newblock \emph{Chemical Communications}, 54\penalty0 (5):\penalty0 427--444,
  2018.

\bibitem[Vroylandt et~al.(2020)Vroylandt, Esposito, and
  Verley]{vroylandt_efficiency_2020}
Hadrien Vroylandt, Massimiliano Esposito, and Gatien Verley.
\newblock Efficiency {Fluctuations} of {Stochastic} {Machines} {Undergoing} a
  {Phase} {Transition}.
\newblock \emph{Physical Review Letters}, 124\penalty0 (25):\penalty0 250603,
  2020.

\bibitem[Arnold(1984)]{arnold_catastrophe_1984}
Vladimir~Igorevich Arnold.
\newblock \emph{Catastrophe {Theory}}.
\newblock Springer Berlin Heidelberg, Berlin, Heidelberg, 1984.
\newblock ISBN 9783540128595 9783642967993.

\bibitem[Poston and Stewart(1996)]{poston_catastrophe_1996}
T.~Poston and Ian Stewart.
\newblock \emph{Catastrophe theory and its applications}.
\newblock Dover Publications, Mineola, N.Y, 1996.
\newblock ISBN 9780486692715.

\bibitem[Duffing(1918)]{duffing1918erzwungene}
Georg Duffing.
\newblock \emph{Erzwungene schwingungen bei ver{\"a}nderlicher eigenfrequenz
  und ihre technische bedeutung}.
\newblock Number 41--42. F. Vieweg \& sohn, 1918.

\bibitem[kor(2008)]{korsch_duffing_2008}
The {Duffing} {Oscillator}.
\newblock In Hans~Jürgen Korsch, Hans-Jörg Jodl, and Timo Hartmann, editors,
  \emph{Chaos: {A} {Program} {Collection} for the {PC}}, pages 157--184.
  Springer, Berlin, Heidelberg, 2008.
\newblock ISBN 9783540748670.

\bibitem[Avetisov et~al.(2019)Avetisov, Markina, and
  Valov]{avetisov2019oligomeric}
Vladik~A Avetisov, Anastasia~A Markina, and Alexander~F Valov.
\newblock Oligomeric “catastrophe machines” with thermally activated
  bistability and stochastic resonance.
\newblock \emph{The Journal of Physical Chemistry Letters}, 10\penalty0
  (17):\penalty0 5189--5192, 2019.

\bibitem[Markina et~al.(2020)Markina, Muratov, Petrovskyy, and
  Avetisov]{markina2020detection}
Anastasia Markina, Alexander Muratov, Vladislav Petrovskyy, and Vladik
  Avetisov.
\newblock Detection of single molecules using stochastic resonance of bistable
  oligomers.
\newblock \emph{Nanomaterials}, 10\penalty0 (12):\penalty0 2519, 2020.

\bibitem[Avetisov et~al.(2021)Avetisov, Frolkina, Markina, Muratov, and
  Petrovskii]{avetisov2021short}
Vladik~A. Avetisov, Maria~A. Frolkina, Anastasia~A. Markina, Alexander~D.
  Muratov, and Vladislav~S. Petrovskii.
\newblock Short pyridine-furan springs exhibit bistable dynamics of duffing
  oscillators.
\newblock \emph{Nanomaterials}, 11\penalty0 (12), 2021.

\bibitem[Kramers(1940)]{KRAMERS1940284}
H.A. Kramers.
\newblock Brownian motion in a field of force and the diffusion model of
  chemical reactions.
\newblock \emph{Physica}, 7\penalty0 (4):\penalty0 284--304, 1940.

\bibitem[Benzi et~al.(1981)Benzi, Sutera, and Vulpiani]{benzi_mechanism_1981}
R~Benzi, A~Sutera, and A~Vulpiani.
\newblock The mechanism of stochastic resonance.
\newblock \emph{Journal of Physics A: Mathematical and General}, 14\penalty0
  (11):\penalty0 L453--L457, 1981.

\bibitem[Benzi et~al.(1982)Benzi, Parisi, Sutera, and
  Vulpiani]{benzi_stochastic_1982}
Roberto Benzi, Giorgio Parisi, Alfonso Sutera, and Angelo Vulpiani.
\newblock Stochastic resonance in climatic change.
\newblock \emph{Tellus}, 34\penalty0 (1):\penalty0 10--15, 1982.

\bibitem[Benzi et~al.(1983)Benzi, Parisi, Sutera, and
  Vulpiani]{benzi_theory_1983}
Roberto Benzi, Giorgio Parisi, Alfonso Sutera, and Angelo Vulpiani.
\newblock A {Theory} of {Stochastic} {Resonance} in {Climatic} {Change}.
\newblock \emph{SIAM Journal on Applied Mathematics}, 43\penalty0 (3):\penalty0
  565--578, 1983.

\bibitem[Gammaitoni et~al.(1998)Gammaitoni, Hänggi, Jung, and
  Marchesoni]{gammaitoni_stochastic_1998}
Luca Gammaitoni, Peter Hänggi, Peter Jung, and Fabio Marchesoni.
\newblock Stochastic resonance.
\newblock \emph{Reviews of Modern Physics}, 70\penalty0 (1):\penalty0 223--287,
  1998.

\bibitem[Wellens et~al.(2004)Wellens, Shatokhin, and
  Buchleitner]{wellens_stochastic_2004}
Thomas Wellens, Vyacheslav Shatokhin, and Andreas Buchleitner.
\newblock Stochastic resonance.
\newblock \emph{Reports on Progress in Physics}, 67\penalty0 (1):\penalty0
  45--105, 2004.

\bibitem[Baughman et~al.(1999)Baughman, Cui, Zakhidov, Iqbal, Barisci, Spinks,
  Wallace, Mazzoldi, De~Rossi, Rinzler, et~al.]{baughman1999carbon}
Ray~H Baughman, Changxing Cui, Anvar~A Zakhidov, Zafar Iqbal, Joseph~N Barisci,
  Geoff~M Spinks, Gordon~G Wallace, Alberto Mazzoldi, Danilo De~Rossi, Andrew~G
  Rinzler, et~al.
\newblock Carbon nanotube actuators.
\newblock \emph{Science}, 284\penalty0 (5418):\penalty0 1340--1344, 1999.

\bibitem[Fujii et~al.(2017)Fujii, Setiadi, Kuwahara, and
  Akai-Kasaya]{fujii2017single}
Hayato Fujii, Agung Setiadi, Yuji Kuwahara, and Megumi Akai-Kasaya.
\newblock Single walled carbon nanotube-based stochastic resonance device with
  molecular self-noise source.
\newblock \emph{Applied Physics Letters}, 111\penalty0 (13):\penalty0 133501,
  2017.

\bibitem[Huang et~al.(2019)Huang, Zhang, Li, and Li]{huang_nonlocal_2019}
Kun Huang, Shuzhu Zhang, Jinhai Li, and Ze~Li.
\newblock Nonlocal nonlinear model of {Bernoulli}–{Euler} nanobeam with small
  initial curvature and its application to single-walled carbon nanotubes.
\newblock \emph{Microsystem Technologies}, 25\penalty0 (11):\penalty0
  4303--4310, 2019.

\bibitem[Liang et~al.(2012)Liang, Huang, Li, Huang, Wu, Fang, Oh, Kozlov, Ma,
  Li, et~al.]{liang2012electromechanical}
Jiajie Liang, Lu~Huang, Na~Li, Yi~Huang, Yingpeng Wu, Shaoli Fang, Jiyoung Oh,
  Mikhail Kozlov, Yanfeng Ma, Feifei Li, et~al.
\newblock Electromechanical actuator with controllable motion, fast response
  rate, and high-frequency resonance based on graphene and polydiacetylene.
\newblock \emph{ACS nano}, 6\penalty0 (5):\penalty0 4508--4519, 2012.

\bibitem[Forns et~al.(2011)Forns, de~Lorenzo, Manosas, Hayashi, Huguet, and
  Ritort]{forns2011improving}
Nuria Forns, Sara de~Lorenzo, Maria Manosas, Kumiko Hayashi, Josep~Maria
  Huguet, and Felix Ritort.
\newblock Improving signal/noise resolution in single-molecule experiments
  using molecular constructs with short handles.
\newblock \emph{Biophysical Journal}, 100\penalty0 (7):\penalty0 1765--1774,
  2011.

\bibitem[Hayashi et~al.(2012)Hayashi, de~Lorenzo, Manosas, Huguet, and
  Ritort]{hayashi2012single}
K~Hayashi, S~de~Lorenzo, M~Manosas, JM~Huguet, and F~Ritort.
\newblock Single-molecule stochastic resonance.
\newblock \emph{Physical Review X}, 2\penalty0 (3):\penalty0 031012, 2012.

\bibitem[Cecconi et~al.(2005)Cecconi, Shank, Bustamante, and
  Marqusee]{cecconi2005direct}
Ciro Cecconi, Elizabeth~A Shank, Carlos Bustamante, and Susan Marqusee.
\newblock Direct observation of the three-state folding of a single protein
  molecule.
\newblock \emph{Science}, 309\penalty0 (5743):\penalty0 2057--2060, 2005.

\bibitem[Lai and Leng(2016)]{LAI201660}
{Zhi-hui} Lai and {Yong-gang} Leng.
\newblock Weak-signal detection based on the stochastic resonance of bistable
  duffing oscillator and its application in incipient fault diagnosis.
\newblock \emph{Mechanical Systems and Signal Processing}, 81:\penalty0 60--74,
  2016.

\bibitem[Lu et~al.(2021)Lu, Wu, Ding, and Chen]{LU2021249}
Ze-Qi Lu, Dao Wu, Hu~Ding, and Li-Qun Chen.
\newblock Vibration isolation and energy harvesting integrated in a stewart
  platform with high static and low dynamic stiffness.
\newblock \emph{Applied Mathematical Modelling}, 89:\penalty0 249--267, 2021.

\bibitem[{Alan Jones} et~al.(1996){Alan Jones}, Karatza, Voro, Civeir, Franck,
  Ozturk, Seaman, Whitmore, and Williamson]{ALANJONES19968707}
R.~{Alan Jones}, Marielena Karatza, Tevita~N. Voro, Pervin~U. Civeir, Annete
  Franck, Orhan Ozturk, John~P. Seaman, Alexander~P. Whitmore, and David~J.
  Williamson.
\newblock Extended heterocyclic systems 1. the synthesis and characterisation
  of pyrrolylpyridines, alternating pyrrole: Pyridine oligomers and polymers,
  and related systems.
\newblock \emph{Tetrahedron}, 52\penalty0 (26):\penalty0 8707--8724, 1996.

\bibitem[{Alan Jones} and Civcir(1997)]{jones1997extended}
R.~{Alan Jones} and Pervin~U. Civcir.
\newblock Extended heterocyclic systems 2. the synthesis and characterisation
  of (2-furyl) pyridines,(2-thienyl) pyridines, and furan-pyridine and
  thiophene-pyridine oligomers.
\newblock \emph{Tetrahedron}, 53\penalty0 (34):\penalty0 11529--11540, 1997.

\bibitem[Sahu et~al.(2015)Sahu, Gupta, Gaur, and Panda]{sahu2015}
Harikrishna Sahu, Shashwat Gupta, Priyank Gaur, and Aditya~N. Panda.
\newblock Structure and optoelectronic properties of helical pyridine–furan,
  pyridine–pyrrole and pyridine–thiophene oligomers.
\newblock \emph{Physical Chemistry Chemical Physics}, 17\penalty0
  (32):\penalty0 20647--20657, 2015.

\bibitem[Abraham et~al.(2015)Abraham, Murtola, Schulz, P{\'a}ll, Smith, Hess,
  and Lindahl]{abr2015}
Mark~James Abraham, Teemu Murtola, Roland Schulz, Szil{\'a}rd P{\'a}ll,
  Jeremy~C Smith, Berk Hess, and Erik Lindahl.
\newblock Gromacs: High performance molecular simulations through multi-level
  parallelism from laptops to supercomputers.
\newblock \emph{SoftwareX}, 1:\penalty0 19--25, 2015.

\bibitem[Kaminski et~al.(2001)Kaminski, Friesner, Tirado-Rives, and
  Jorgensen]{kam2001}
George~A Kaminski, Richard~A Friesner, Julian Tirado-Rives, and William~L
  Jorgensen.
\newblock Evaluation and reparametrization of the opls-aa force field for
  proteins via comparison with accurate quantum chemical calculations on
  peptides.
\newblock \emph{The Journal of Physical Chemistry B}, 105\penalty0
  (28):\penalty0 6474--6487, 2001.

\bibitem[Bussi et~al.(2007)Bussi, Donadio, and Parrinello]{bussi2007canonical}
Giovanni Bussi, Davide Donadio, and Michele Parrinello.
\newblock Canonical sampling through velocity rescaling.
\newblock \emph{The Journal of Chemical Physics}, 126\penalty0 (1):\penalty0
  014101, 2007.

\end{thebibliography}

\end{document}


\section{Simulation protocol}
\subsection{Parameters for Molecular Dynamics simulation\label{SI1}}
Morphology simulations were performed using the Gromacs2019 \cite{abr2015} simulation package. Lennard– Jones parameters were taken from the OPLS-AA\cite{kam2001} force field with a scaling factor of $0.5$ for the $1–4$ interactions. Long-range electrostatic interactions were treated using a smooth particle mesh Ewald technique\cite{ess1995} with a cut-off of \SI{1.2}{\nano\metre}. Bond vibrations were constrained with a LINCS \cite{hes2008} algorithm. All calculations were performed in the NVT ensemble using the canonical velocity-rescaling thermostat, as implemented in the Gromacs2019 simulation package.

\begin{figure}[H]
    \includegraphics[width=\linewidth]{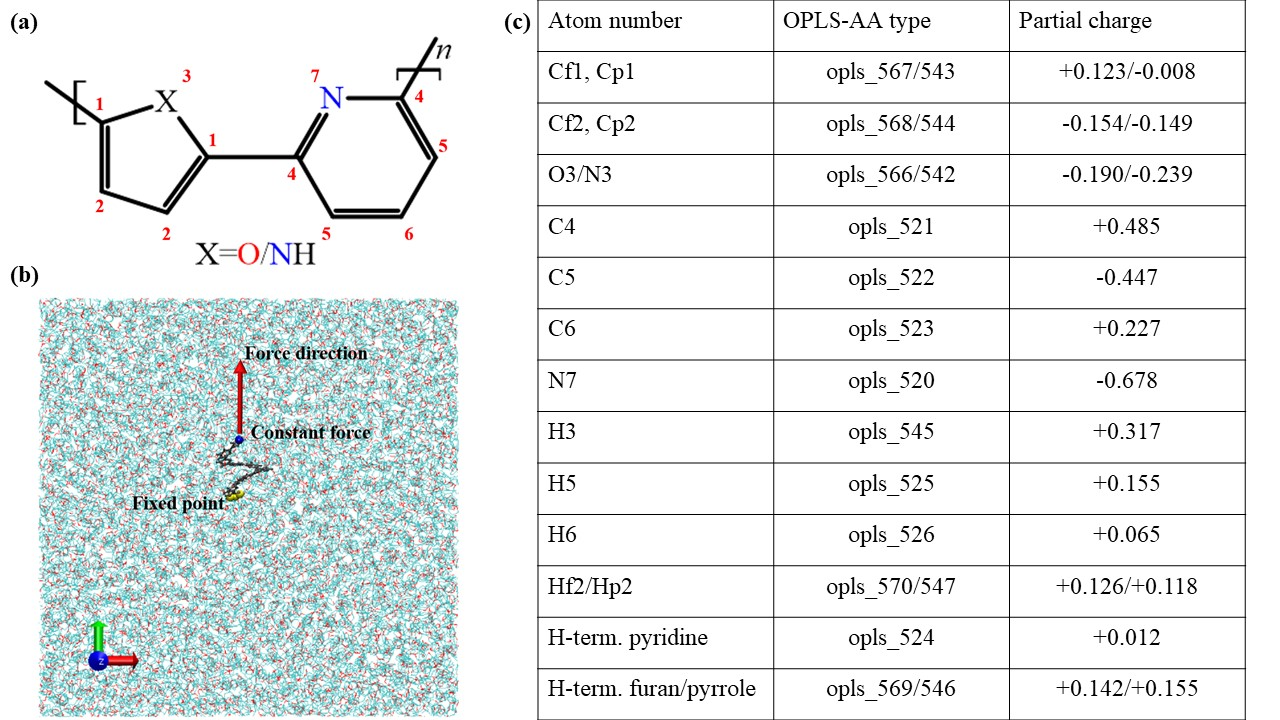}
    \centering
\caption{(a) The chemical structure of a PF/PP monomer, (b)a scheme of the simulation box for the oligo-PF5 system, (c) parameterization of PF/PP oligomers in OPLS-AA force field types and partial charges.}
\label{fig:S1}
\end{figure}

The parameterization of oligo-PP springs is presented in Figure \ref{fig:S1}a-b. The OPLS-AA force field contains parameters only for single pyridine and furan molecules. First, $C4$ and $C6$ atoms have a covalent bond instead of hydrogen atoms and the partial charge of hydrogens were added to carbon atoms to maintain the neutrality. All the other partial charges are taken from OPLS-AA force-field. Second, there is only one undefined angle bond for atoms $C4-C6-N11$, which is $CW-CA-NC$ angle type in OPLS-AA. We used $CA-CA-NC$ angle due to similarity of CA and CW types. Both correspond to aromatic $sp2$-hybridized carbon.

The simulation was started from a random initial configuration. An equilibrated state were reached in two steps. First, a short run of \SI{10}{\pico\second} was carried out in the NVT ensemble with a time step of \SI{0.01}{\femto\second} with the leapfrog integrator for motion equations. Then the second part of equilibration was done for another \SI{10}{\nano\second} with a time step of \SI{2}{\femto\second}. After the equilibration, a long trajectory of \SI{500}{\nano\second} is obtained to achieve a clear picture of transitional behaviour.  

To simulate the behaviour of the oligo-PP$5$ under the external load, one end of it was fixed and the other end was pulled by an external force (see Figure \ref{fig:S1}c). We used a simulation box of size $7\times7\times7$ $nm^3$ due to technical aspects of Gromacs pulling algorithm. The axis of an initial conformation of oligo-PP$5$ was oriented in the XY plane. The longitudinal load F was applied along Z-direction to the center of mass of the last monomer unit and directed toward the attraction point, which was located along the vector connecting the left and right ends of the molecule.

In both cases, the center of mass of the first monomeric unit was fixed using a spring potential of $k=\SI{100}{\kilo\joule}\cdot\SI{}{\per\mole}\cdot\SI{}{\per\nano\metre\squared}$; other specific constraints for bond length or atom positions were applied.
 
\subsection{Oligo-PP-$5$ in water}
In order to find out the impact of THF on the dynamics of oligo-PP, we also model its stretching in water. The simulation protocol in this case is similar to the one described in the section Parameters for Molecular Dynamics simulation above. One end of the oligo-PP$5$ was fixed and the pulling force was applied to another end. The end-to-end ($R_{e}$) distance both in water gradually increases along with the force up to the value of $F$ = \SI{305}{\pico\newton} at which point the oligo-PP$5$ helix turns to the fully stretched state
. Note that the oligo-PP$5$ in water is present in the squeezed state at the pulling forces below $F$ = \SI{305}{\pico\newton} due to the hydrophilic nature of the solvent.


\subsection{Parameters for periodic signal}\label{SI:2}
Stochastic resonance was obtained by applying a periodic signal leading to the swinging of the oligo-PF/PP bistable potential. An oscillating force was implemented by setting a charge ($-1$) on the oligomer, adding a compensating charge ($+1$) as a counter ion in the solvent and applying a periodical electric field. The additional charge was placed on the end group of the oligo-PF-$5$/oligo-PP-$5$ and spread between the first, second and fifth atoms of the furan/pyrrole group (see Fig. \ref{fig:S1}a). The additional charges are equally distributed between the chosen atoms.

The periodic field in the Gromacs2019 package is defined by an equation $$ E(t) = E_0 \exp{\left[-\frac{(t-t_0)^2}{2\sigma^2}\right]}\cos{\omega(t-t_0)},$$ where the exponential part modulates the periodic part with the pulsing behaviour, $E_0$ is the amplitude of the signal and $\omega t_0$ is the oscillation phase. Here we use only the static part when $\sigma=0$ along y-axes (see Figure \ref{fig:S1}b). An external oscillating electrical field was directed along the constant force direction in the oligo-PF-$5$/oligo-PP-$5$ case.

\subsection{Influece of the charges}\label{SI:3}

The SR regime of the both oligo-PP-$5$ and oligo-PF-$5$ springs was examined by applying an additional oscillating electrical force on the unit charge preset on the pulling end of the spring while a compensative charge was placed $\sim \SI{2.2}{\nano\metre}$ "below" the spring (if pulling direction is "up"). However, such a modification changed drastically the bistable behaviour of the springs: even with $\vec{F} = 0$ the bistability was observed with spontaneous vibrations present, while the equilibrium value of $F_{eq}$ (when mean time spent in open and closed states is approximately equal) for oligo-PP-$5$ was only $  \SI{3.5}{\pico\newton} $. This result indicates that an artificial negative charge decreases the size of the closed state energy minimum, presumably because of its interactions with partial charges on the aromatic ring that comes in contact with the charged end of the spring in closed state. We also observed that positively charged end of the spring influences the $F_{eq}$ in opposite way (it increases compared to the non-charged case), which partially confirms the hypothesis about interactions with partial charges.

\section{Results}

\subsection{Mean lifetime estimation}
\begin{figure}[H]
    \includegraphics[width=\linewidth]{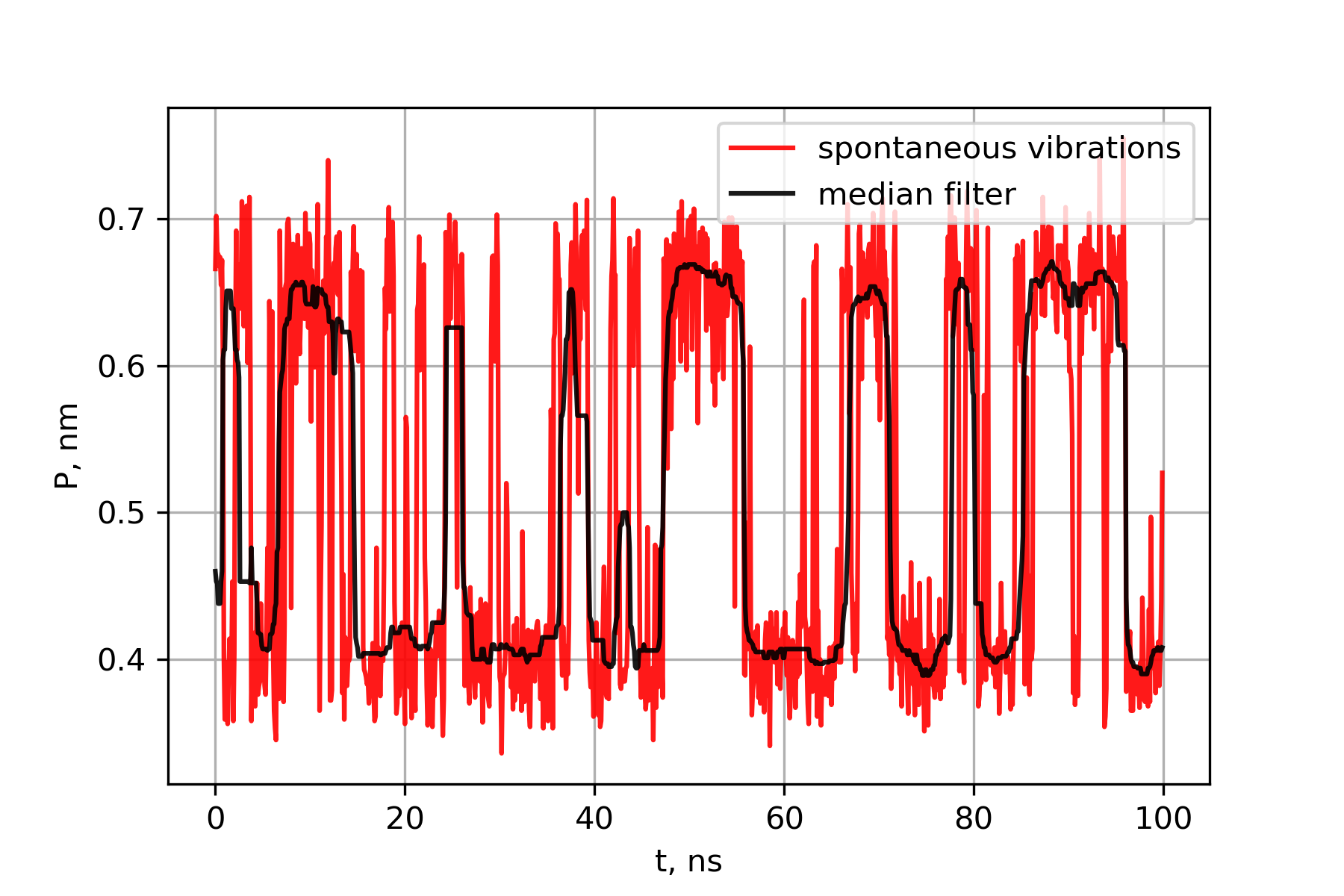}
    \centering
\caption{ The spontaneous vibrations trajectory with and without the median filter usage.}
\label{fig:S5}
\end{figure}
In the oligo-PF/PP systems, one can observe the dynamics at two temporal scales: local and global. The first one is characterized by fast fluctuations near the squeezed/stress-strain states in case. However, the global dynamics (the switching between the ends) is more important for stochastic resonance. To calculate the mean lifetime, the local dynamics were excluded using the median filter $ndimage.median\_filter$ from an open-source software SciPy for Python 3 with a median filter window size of \SI{3.5}{\nano\second}, which corresponds to local vibrations. The resulting trajectory is shown in \ref{fig:S5}. According to the estimation, the mean lifetime for oligo-PF$5$ was $\tau=\SI{6.14}{\nano\second}$ and corresponded to the external force $F=\SI{279}{\pico\newton}$.

\bibliography{references}